\documentstyle[preprint,revtex]{aps}
\begin{document}
\draft
\date{10 May 1992}

\begin{title}
 PRESSURE INDUCED TOPOLOGICAL PHASE TRANSITIONS IN MEMBRANES
\end{title}
\author{P.T.C. So\cite{*}, Sol M. Gruner and
Shyamsunder Erramilli \cite{footnote}}
\begin{instit}
Department of Physics, Joseph Henry Laboratories
Princeton University
P.O. Box 708
Princeton, NJ  08544
\end{instit}

\begin{abstract}

Some highly unusual features of a lipid-water liquid crystal are
revealed by high pressure  x-ray diffraction,  light scattering and
dilatometric studies  of the  lamellar (bilayer  $L_{\alpha}$)  to
nonlamellar inverse  hexagonal ($H_{II}$)  phase transition.
 (i) The  size of the unit cell of the $H_{II}$
phase   increases with  increasing pressure.  (ii)  The
transition volume,  $\Delta  V_{bh}$,  decreases  and  appears  to
vanish as  the pressure is increased. (iii) The intensity of scattered light
increases  as  $\Delta V_{bh}$
decreases.  Data  are  presented  which  suggest  that  this
increase is  due to  the formation  of an intermediate cubic
phase, as predicted by recent theoretical suggestions of the
underlying universal phase sequence.

\end{abstract}

\narrowtext

     Dispersions of  biological lipids  in water  exhibit  a
rich  variety   of  structurally   complex  phases   as  the
temperature, pressure  and  water  content  are  varied.
The  lamellar liquid  crystalline bilayer ($L_{\alpha}$) to
inverse  hexagonal  ($H_{II}$)  phase  transition  has  attracted
particular interest  because  it  is  the  simplest  of  the
lyotropic  transitions   in  which  both  the  topology  and
curvature of  the lipid-water  interfaces  internal  to  the
liquid crystal  change  discontinuously  at  the  transition
\cite{dope_references}.

The presence of water and
electrostatic  effects   makes  lyotropic  dispersions  very
complicated  and  difficult  to  understand.
Insight into   understanding  such
systems first came with the recognition  that they form assemblies of
2-dimensional liquid-like  films  which  elastically  resist
lateral extension  or compression,  or bending  away from  a
desired curvature.
Winsor \cite{winsor71}  and Helfrich  \cite{helfrich73} suggested
that much  of the  complexity of  the molecular interactions
can be  lumped into  phenomenological parameters such as the
 curvature  of the lipid-water interfaces and the
elastic moduli  of the  lipid layers.
The  utility   of  this  phenomenological approach  toward
understanding   the    lamellar   bilayer   to   nonlamellar
transitions has  been established  by Gruner  and  coworkers
\cite{kirk84} and  by Charvolin  \cite{charvolin85}.

     The idea that many aspects of the global phase behavior
can be  explained without considering the detailed molecular
structure  of  the  molecules  involved  led  Goldstein  and
Leibler \cite{goldstein88}  to propose  a mean-field  model for  the  well
known bilayer chain melt (i.e. gel-liquid crystalline) phase
transition. They  proposed a  Ginzburg-Landau approach using
the strain  field of the lipid layers, with phenomenological
parameters modulated  by  microscopic  interactions  arising
from van  der Waals,  hydration  and  electrostatic  forces.
While this  approach has  not been  extended to  nonlamellar
phases, the  enormous simplification  afforded by Helfrich's
insight has  stimulated recent  experimental \cite{dope_references} and
theoretical interest  in  nonlamellar  phases  \cite{anderson88,chen90}.
A rather general  attempt in  this direction  is  the  Landau-
Ginzburg  model  of  Chen  et  al  \cite{chen90},  which  involves  a
combination of  scalar and  vector order  parameters that is
claimed to  have, in  principle,  sufficient  complexity  to
describe not  only lamellar  and hexagonal  phases, but also
more geometrically  complex  phases,  such  as  bicontinuous
cubic   phases  \cite{anderson88,hyde84}.  However,  the  large  number  of
phenomenological parameters  involved, as  well as a lack of
an intuitive  physical interpretation  or ready experimental
handle on the order parameters, suggests that deeper insight
may be derived from simpler explanations of unusual phase or
structural behavior  in cases  where care  has been taken to
limit the  energetic significance  of complicating  factors,
such as  the effects  of  electrostatic  charge  or  limited
water. Although there
have been  numerous studies  of the  effects of variation of
temperature on  these transitions  \cite{dope_references}, there have been very
few investigations  of the  effects of pressure.
In  this Letter,  we describe  some rather unusual phenomena
exhibited  by   excess-water  dispersions   of  electrically
neutral lipid, 1,2-dioleoyl-sn-glycero-3-phosphoethanolamine
(DOPE; also known as dioleoylphosphatidylethanolamine), which swells
in water to form phases with a well-defined water fraction,
with a view to
stimulating quantitative models.

     1) {\it Nonlamellar  phase transitions temperatures are very
pressure sensitive.}   Optical  scattering    \cite{yeager83}
and x-ray diffraction \cite{shyamsunder89} measurements suggested
that the  $L_{\alpha}$-$H_{II}$ transition temperature may be depressed
by  surprisingly   modest  pressures. Our
experiments were conducted
using a  rotating anode x-ray source, a point-focussed beamline, a
thermostated high  pressure beryllium x-ray cell, and a two-
dimensional  x-ray  detector  \cite{so92}  which,  in  combination,
allowed rapid  surveys of the temperature-pressure structure
of small ($\sim 10$ mg) lipid samples. The phases and internal
dimensions of  the lipid-water  liquid crystals were readily
identified by  the characteristic x-ray diffraction patterns
of  unoriented   dispersions  \cite{luzzati68}.   The  Clausius-Clapeyron
coefficient  of   change  of   transition  temperature  with
pressure was  $dT_{bh}/dP=44$ K/kbar,  a value  which  is  nearly
twice as  large as for the chain melting transition \cite{stamatoff78} and
which establishes  the excess-water $L_{\alpha}$-$H_{II}$ transition as
one of  the most  pressure sensitive first order transitions
known  in   a  non-gaseous   system.  The   isothermal  bulk
compressibility of  the lipid-water  system is comparable to
that of water (see below). Since the system is not unusually
compressible,  this   suggests  that  the  $L_{\alpha}$-$H_{II}$  phase
transition involves  a very  delicate balance  of  competing
energetic contributions  such that  small changes in overall
volume result in large changes in the balance.

     2) The  unit cell  spacing compressibility is negative.
Figure 1  shows that  the unit  cell spacing,  $b$, of the $H_{II}$
phase increases  as pressure  is applied.  Furthermore, this
pressure  dependence   is  remarkably  steep:  at  20  C,  $b$
increases by nearly 1 \AA  for every 100 bar increase in pressure. The
increase in  the unit cell is almost all due to the transfer
of water  from the  coexisting  water  bulk  to  the  liquid
crystal \cite{narayan90}.  The data yield a value of the isothermal bulk
linear compressibility  of the  Bragg  lattice  of  $\alpha_{T}=-
\frac{1}{b}\left( \frac{\partial b}{\partial  P} \right)_{T}=-2.5
\times 10^{-2}$ bar$^{-1}$. Not only is $\alpha_T$ negative, but it is
more than 1500 times as large as the compressibility of water.
We define  a
generalized           Clausius-Clapeyron           relation,
$\frac{dT}{dP} =
\frac{\Delta T}{\Delta \xi}\frac{\Delta \xi}{\Delta P}$,       that
relates the  sensitivity of a structural parameter, $\xi$,
to   pressure    and   temperature    changes.    We    find for the
unit cell spacing with $\xi = b$,
$\Delta b/\Delta T= 0.22$  A/K and  $\Delta b/\Delta P = 9.6$
\AA/kbar, giving for  $\frac{dT}{dP} = 44$  K/bar, which is essentially
the same as the Clausius-Clapeyron  value for  the $L_{\alpha}$-$H_{II}$
phase  transition   given  above.  {\it This  suggests  that  the
thermodynamics of  the phase  transition may  be  understood
using models based only on the gross structural parameters, without
reference to details of the molecular interactions.}

 Thermodynamic stability arguments show that the
apparent  negative  compressibility  in
Figure 1  refers only  to the increase in the unit cell size
of  the   two-dimensional  hexagonal  lattice;  the  overall
specimen volume  must decrease with  increasing pressure. Almost
all of  the increase in size with increasing pressure arises
from an  increase in the radius of the central water core of
the hexagonally  packed tubes  and the  resultant influx  of
bulk water. Electron density reconstruction of the x-ray data show that
the truly pressure sensitive aspect of the
system is  the spontaneous  curvature  of  the  lipid  layer
making up  the hexagonally  packed tubes \cite{narayan90}.  Measurement of
parameters such as the
average  volume,  length  and  interfacial  area  per  lipid
molecule cannot be obtained from x-ray diffractometry alone; therefore  we
constructed  a  high  pressure  (0 - 3 kbar) dilatometer  capable  of
resolving volume  changes of about 0.1 nl/mg in small lipid-
water specimens \cite{so92b}. Figure 2
shows the  specific volume  as a  function  of  pressure  at
various temperatures.

     3) At  least four  different pressure  sensitive  phase
transitions are  observed in  excess water-DOPE samples (Fig.2 ).
The
isothermal bulk  compressibility of the DOPE-water system is
about $\beta_{T} = -\frac{1}{V}\left(\frac{\partial V}{\partial P}\right)_{T}
= + 2\times 10^{-4}$ bar$^{-1}$.  Phase  transitions are  readily  identified
in  the
dilatometry data  as discontinuities  in the  lipid specific
volume or  as slight  changes in  the slope of the isotherms
signalling a  change in compressibility. Figure 2 shows that
four distinct  transitions  corresponding  to  five  phases, which
have also  been examined via x-ray diffraction \cite{so_thesis}.
(i) At  13 C
 a small volume change occurs at about 200 bar
which corresponds to the $L_{\alpha}$-$H_{II}$ transition.
(ii) At 8 C  and 750  bar, a  large volumetric  change is
observed which  is identified from the appearance of sharp peaks in the
wide angle x-ray scattering
(WAXS)  as   the  familiar  chain  melting  (gel  to  liquid
crystalline) transition.  The volume  change is in agreement
with previous  thermodynamic studies  on lipid  systems with
phosphatidycholine (PC) headgroups \cite{utoh85}.
(iii) At 8 C and 1250 bar, a phase transition with a small
volume change is observed
with the position and
intensity of the sharp peaks in the
WAXS changing discontinuously, suggesting that the symmetry of the
packing of the hydrocarbon chains has been changed.
(iv) A  large volume  change is observed at 8 C and 2 kbar
the system  forms a high-pressure phase characterized by five
 distinct WAXS lines. This suggests
that the  chains are  packed with a high degree of order and
that  the  phase  is  likely  to  be  a  crystalline  phase.
Crystalline phases  corresponding to sub-gel transitions are
known in PC systems \cite{wong84,utoh85}.

     4) The  transition volume, $\Delta V_{bh}$, enthalpy, $\Delta H_{bh}$
and  entropy,   $\Delta S_{bh}$,  of   the  $L_{\alpha}$-$H_{II}$   transition
decreases as  the  pressure  is  increased.  The  transition
volume  was  measured  by  dilatometry  and  the  transition
enthalpy and  entropy were  then computed using the Clausius-Clapeyron
relation.
 The
transition volume  decreases as  the  temperature  increases
(inset,  Figure   2).  Most   interestingly,  $\Delta V_{bh}$,  and
therefore $\Delta H_{bh}$ and $\Delta S_{bh}$, become too small to measure
with our apparatus at about 80 C and $ \sim 1750$ bar.

     5) Light  scattering of the DOPE-water system increased
as the  transition volume  decreased.
 A conventional high
pressure optical  cell was  constructed  out  of  beryllium-
copper and  equipped with  sapphire windows  and unsupported
area seals.  The intensity,  $I(q)$, of  scattered He-Ne laser
light ($\lambda =632$  nm) was  measured with  a light  sensitive
diode at an angle corresponding to  $q  \sim 10^{-3}$ nm$^{-1}$.
 A plot of $I(q)$ shows a prominent
peak at  a temperature   in  agreement with
the $L_{\alpha}$-$H_{II}$ transition, as identified by both dilatometry
and x-ray diffraction.

     6) A  cubic phase  occurs between  the $L_{\alpha}$  and  $H_{II}$
phases.
It might be imagined that the  increase in  intensity of  the scattered  light
concomitant  with   a  decrease   in  transition  volume  is
suggestive of  some type  of critical  behavior.  However, a different
hypothesis provides a more plausible explanation:
the light-scattering
anomalies observed in the L$_{\alpha} $ - H$_{\rm II} $
transition are due to the onset of an intermediate phase, associated
with very little latent heat, but whose formation is kinetically
hindered by large activation barriers.
 Defects forming the precursor
to the intermediate phase scatter light.

     In order  to test this hypothesis, an excess water-DOPE
specimen was  allowed to equilibrate at $80 \pm 0.05$ C and $1750
\pm 3$  bar,  the  nominal  `critical  point'.  In  situ  x-ray
diffraction patterns  of the sample were taken roughly every
hour for three days. The initial diffraction pattern (Figure
4a) is  characteristic  of  the  hexagonal  phase.  As  time
elapses, diffuse  scatter becomes  prominent
concomitant  with   a  decrease  in  the  intensity  of  the
hexagonal diffraction  peaks. Within  12  hours,  additional
peaks appear. After 36 hours, the diffraction pattern can be
readily indexed to a cubic lattice of $Pn3m/Pn3$ symmetry.
     The x-ray  data shows  that a cubic phase forms at high
temperature and  pressure. The  spontaneous formation of the
cubic phase  at the expense of the hexagonal phase indicates
that the  cubic phase  is stable.  The slow formation of the
cubic phase  and the  coexistence seen  in Figure 4 with the
metastable hexagonal  phase are  indicative of  a pronounced
kinetic barrier between the phases.

 The sequence of phase dimensionality,
from 1-D lamellar to 3-D cubic to 2-D hexagonal, underscores
the  fact   that  the  lattice  dimensionality  is  not  a relevant
parameter for these topological phase transitions.
Gruner \cite{kirk84}, and
Charvolin \cite{charvolin85}  suggest that the phase behaviour can be
understood as a result of a {\it frustration} between opposing requirements of
membranes that have to curl to satisfy curvature requirements on the one hand
and to try to maintain constant bilayer thickness on the other. Because
the phases are formed by a subtle balance of competing interactions,
the latent heat associated with the transitions can be small.
Anderson  et   al  \cite{anderson88}   argued   that   certain
bicontinuous cubic  phases provide  equilibrium  compromises
between lamellar  and inverse hexagonal phases such that one
expects them  to always  occur between  these phases  in the
phase diagrams. The data presented in this Letter
 support this  picture,
and suggest  that the  underlying universality of this phase
sequence   may be  obscured by  the large kinetic barriers.
The origin of the large barriers lies in the fact that
the  $L_{\alpha}$, $H_{II}$, and bicontinuous cubic phases have
very different    topologies  of   the  interfaces.
Transitions between  lamellar and non-lamellar phases
necessarily require lipid-water interfaces to be
torn and reformed.  The process requires exposing the
hydrocarbon tails of the lipid molecules to water.
The energy associated with  this disruption probably is the
large activation barrier.

Rather striking  support
for  the
notion of an underlying universality comes from the observation
diblock copolymers exhibit essentially the same sequence of phases
\cite{grunerphyschem}, although the microscopic interactions are quite
different.
It
would be interesting to see if the observed negative anisotropic
compressibility also carries over to the diblock copolymer systems.
More importantly,
a unified quatitative theory of all systems exhibiting topological phase
transitions may now be possible.

\section{ACKNOWLEDGEMENTS}

This work was supported by the ONR (Contract N00014-86-K-0396 P0001), and
by the DOE (Grant DE-FG02-87ER60522) and by the NIH (Grant GM32614).  We
thank R. Bruinsma, J. T. Gleeson,
D. C. Turner and M. W.
Tate  for useful discussions and assistance with figures.

\figure  {\it Unit cell spacing in the $H_{II}$ phase as a
function of pressure at various temperatures.}  Data are shown
for aqueous dispersions of high purity DOPE purchased from Avanti Polar
Lipids at the temperatures 25 C ($\bigcirc$), 43 C ($\bigtriangledown$),
62 C ($\bigtriangleup$) and 81 C ($\Box$).
Sample purity was periodically checked
before and after experiments by thin layer chromatography (TLC).
Lyophilized lipid was
mechanically mixed with excess
deionized water and thermally cycled several times between -10$^\circ$C and
20$^\circ$C to ensure mixing.  The hydrated lipid was transferred to a
high-pressure  cell designed for x-ray measurements. The dashed line indicates
the approximate phase boundary between the $H_{II}$ and $L_{\alpha}$ phases.

\figure {\it $P-V$ isotherms  for the DOPE-water
system.}  Solid lines indicate experimental data. Isotherms at 5$^{0}$ C
intervals from
8 to 83
$^{0} $C (from left to right) are shown. The phase transitions
indicated are (i) the inverted
hexagonal to lamellar transition $H_{II} - L_{\alpha}$; (ii)
the chain melting
transition $L_{\alpha}-L_{\beta}$; (iii) and (iv) are gel-gel phase transitions
not previously reported in the
 DOPE-water system, but are presumed
to be similar to reported transitions in other lipid-water
systems \cite{wong84}. Inset shows that the transition volume  $\Delta V_{bh}$
for the $L_{\alpha}-H_{II}$ transition derived from the isotherms
 using a procedure
similar to that of Wilkinson and Nagle \cite{wilkinson78}. $\Delta V_{bh}$
decreases as function of pressure and becomes too small
to measure as the pressure reaches $\sim 1750$ bar.

\figure {\it Light scattering study of the DOPE-water system.} Data
were taken at a fixed value of the magnitude of the scattering vector ($q =
10^{-3}$ nm$^{-1}$), at the pressures indicated.
The increase in light scattering intensity coincides with the decrease in
$\Delta V_{bh}$.

\figure {\it X-ray diffraction studies at the point at which $\Delta V_{bh}$
becomes too small to measure.}
Evolution of the x-ray diffraction pattern at the times indicated after
reaching 1750 bar, 80 C.
The locations of the Bragg peaks in the
diffraction pattern is characteristic of an inverted
hexagonal phase (indexed in curly brackets).  As time elapses, additional Bragg
peaks emerge that index to a
cubic lattice (square brackets).
The best fit to the observed Bragg peaks corresponds to a
lattice of $Pn3m$ (or $Pn3$) symmetry shown schematically in the lower left.
that has been observed in other
lipid-water systems. The  cubic phase may also be
obtained by
repeatedly thermally cycling the system across the $L_{\alpha}-H_{II}$
phase transition \cite{shyamsunder88}.


\begin{references}
\bibitem{footnote} To whom correspondence should be addressed.
\end{references}
\end{document}